\documentclass[aps,prl,twocolumn,superscriptaddress,showpacs,a4paper,english]{revtex4}
\usepackage{amsfonts,epsfig}
\usepackage{amsmath}
\usepackage{epstopdf}
\usepackage{graphicx}
\usepackage{bm}
\usepackage{color}
\usepackage{amssymb}
\usepackage{ulem}
\usepackage{times}
\usepackage{dcolumn}
\usepackage{cases}
\usepackage{txfonts}
\usepackage[english]{babel}
\usepackage{epstopdf}
\usepackage{hyperref}
\usepackage{soul}
\DeclareGraphicsExtensions{.png,.eps}

\sethlcolor{yellow}

\begin{document}

\title{Criticality in Two-Dimensional Quantum Systems: Tensor Network Approach}

\author{Shi-Ju Ran}
\affiliation{ICFO-Institut de
	Ciencies Fotoniques, The Barcelona Institute of Science and
	Technology, 08860 Castelldefels (Barcelona), Spain}
\author{Cheng Peng}
\affiliation{CAS Key Laboratory for Vacuum Physics, School of Physics, University of Chinese Academy of Sciences, P. O. Box 4588, Beijing 100049, China}
\author{Wei Li}
\affiliation{Department of Physics, Beihang University, Beijing 100191, China}
\author{Maciej Lewenstein} 
\affiliation{ICFO-Institut de
	Ciencies Fotoniques, The Barcelona Institute of Science and Technology, 08860 Castelldefels (Barcelona), Spain}
\affiliation{ICREA-Instituci\'{o} Catalana de Recerca i Estudis Avan\c{c}ats, Lluis Companys 23, 08010 Barcelona, Spain}
\author{Gang Su}
\email[Corresponding author. ]{Email: gsu@ucas.ac.cn}
\affiliation{CAS Key Laboratory for Vacuum Physics, School of Physics, University of Chinese Academy of Sciences, P. O. Box 4588, Beijing 100049, China}

\begin{abstract}
 Determination and characterization of criticality in two-dimensional (2D) quantum many-body systems belong to the most important challenges and problems of quantum physics. In this paper we propose an efficient scheme to solve this problem by utilizing the infinite projected entangled pair state (iPEPS), and tensor network (TN) representations. We show that the criticality of a 2D state is faithfully reproduced by the ground state (dubbed as boundary state) of a one-dimensional effective Hamiltonian constructed from its iPEPS representation. We demonstrate that for a critical state the correlation length and the entanglement spectrum of the boundary state are essentially different from those of a gapped iPEPS. This provides a solid indicator that allows to identify the criticality of the 2D state. Our scheme is verified on the resonating valence bond (RVB) states on kagom\'e and square lattices, where the boundary state of the honeycomb RVB is found to be described by a $c=1$ conformal field theory. We apply our scheme also to the ground state of the spin-1/2 XXZ model on honeycomb lattice, illustrating the difficulties of standard variational TN approaches to study the critical ground states. Our scheme is of high versatility and flexibility, and can be further applied to investigate the quantum criticality in many other phenomena, such as finite-temperature and topological phase transitions.
\end{abstract}

\pacs{03.65.Ud, 71.27.+a, 74.40.Kb}

\maketitle

\textit{Introduction.}--- Considerable efforts have been devoted to explorations of novel properties of two-dimensional (2D) quantum many-body systems. In such systems, rich geometries of 2D lattices, and strong competition between quantum fluctuation and magnetic ordering, provide a fertile ground for various exotic phenomena \cite{ManybodyBook}. Among others, the 2D frustrated Heisenberg models (e.g. the kagom\'{e} antiferromagnet \cite{ExpKagome}) were shown to be good candidates of realizing quantum spin liquids (QSLs) \cite{QSL}. Currently it is still of great interest in condensed matter physics to address elusive properties of the QSLs like topological orders \cite{TopoOrder}, fractional excitations \cite{FractionalE}, and criticality \cite{QPT}, etc.

However, many important issues of 2D systems remain unsolved due to the lack of efficient methods. Even for models with local interactions, the entanglement in ground states increases with the boundary length between the sub-systems (obeying the so-called area law \cite{AreaLaw}), making the model difficult to study. In addition, the artificial gap due to finite-size effect makes it even harder to access the criticality. Unfortunately, most of the recognized many-body algorithms are efficient only for finite-size systems, including quantum Monte Carlo \cite{Critical2DQMC} and density matrix renormalization group \cite{DMRG}. Efficient algorithms for infinite 2D systems are still in urgent demand. 
% There are still no efficient ways to generalize these algorithms for infinite 2D systems.

Tensor network (TN) has been widely accepted as a powerful tool to investigate 2D quantum systems \cite{ReviewTNS}. For example, matrix product states (MPS) \cite{MPSPEPS}, and their higher-dimensional generalization, called projected entangled pair states (PEPS) \cite{MPSPEPS,PEPS} naturally fulfil the area law of entanglement \cite{PEPSCritical}. TN have also been proven successful to construct non-trivial states such as resonating valence bond states (RVB) \cite{PEPSCritical,RVBPEPS}, and string-net states \cite{StringNet}. On the other hand, they provide faithful variational ansatz for non-critical ground-states \cite{PEPS}, and for thermodynamic \cite{ODTNS} simulations. With a powerful ansatz, the following task would be to optimize and extract the physical properties of the system under consideration; unfortunately this is very difficult in 2D.

Optimization task consists essentially in finding an efficient way to contract a TN, which in general can be done only approximately \cite{PEPS,NCD}. Many algorithms have been developed to achieve this task, including tensor renormalization group \cite{TRG1}, time-evolving block decimation \cite{iTEBD}, tensor network encoding schemes \cite{ODTNS,NCD,AOP}, and so on. Even with the implementation of PEPS (by either construction or optimization), it is still very challenging to extract useful physical information. One straightforward way is to compute the average of an operator, e.g. energy or magnetization, which is useful to study the states that obey Landau's paradigm (i.e. they exhibit local order parameters \cite{Landau}). For some exotic states of matter including QSLs, the quantities to characterize their nature may be non-local (e.g. topological orders \cite{TopoOrder}), or even simply unknown. Many efforts have been realized to settle down this issue; in particular boundary theories \cite{BoundaryH0,BoundaryH1,BoundaryH2} provide novel insight into topological orders in 2D.

In this paper, we propose a general scheme to determine the criticality of 2D quantum many-body systems in infinite lattices. By mapping an infinite PEPS (iPEPS) $|\Psi \rangle$ onto a 2D TN simply with $\langle \Psi | \Psi  \rangle$, an effective 1D Hamiltonian $\varrho$ is defined by an infinite stripe of the TN. We rigorously demonstrate that the criticality of an iPEPS can be robustly reproduced by the ground state of $\varrho$ (named the boundary state of the iPEPS). By increasing the bond dimension ($D$) of the boundary state (in the MPS form), its entanglement spectrum varies essentially in two different ways for gapped and critical iPEPS. For a gapped iPEPS, the Schmidt numbers $\{\lambda_i\}$ of the boundary state do not change when the bond dimension $D$ increases. In contrast,  for a critical iPEPS, $\{\lambda_i\}$ are squeezed as $D$ increases, forming a completely different pattern. Consequently, the entanglement entropy converges to a finite value as $D$ increases, when the iPEPS is gapped, or diverges logarithmically, when it is critical. This is consistent with the existing theories of criticality proposed in 1D systems \cite{EntCritic}. We verify our scheme for the RVB states on kagom\'e and honeycomb lattices. These two states can both be written as iPEPS with only $D=3$ \cite{RVBPEPS}, but one is gapped and the other is critical \cite{RVBCritic}. Consequently, the entanglement spectrum of the boundary state faithfully identifies the criticality, where the boundary state of the honeycomb RVB is found to be described by the $c=1$ conformal field theory (CFT) \cite{CFT,CFT_Ent}.

Furthermore, we apply our scheme to the ground-state iPEPS of the XXZ model on honeycomb lattice by the simple update algorithm \cite{TRG2}. The complexity of capturing the criticality of a 2D ground state with variational TN schemes is discussed. Our work provides a reliable way to investigate the criticality of 2D states with boundary states and CFT.

\begin{figure}[tbp]
	\includegraphics[angle=0,width=1\linewidth]{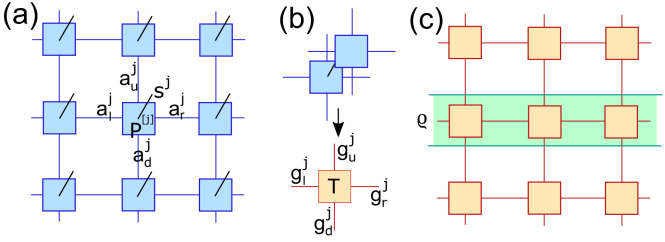}
	\caption{(Color online) (a) The graphic representation of an iPEPS $|\Psi \rangle $ formed by infinite copies of the local tensor $P$ [Eq. (\ref{eq-PEPS})]. (b) The local tensor $T$ [Eq. (\ref{eq-LocalT})] of the TN $\langle \Psi | \Psi \rangle $. (c) The graphic representation of the TN $\langle \Psi | \Psi \rangle $, where the 1D effective Hamiltonian $\varrho$ is defined as an infinite tensor stripe (green shadow).}
	\label{fig-PEPS}
\end{figure}

\textit{Correspondence between a 2D iPEPS and a 1D effective quantum Hamiltonian.}--- In the following, by utilizing TN and MPS, we prove that the criticality of a 2D quantum state written as iPEPS is reproduced by the ground state of a 1D effective Hamiltonian. Taking square lattice as an example, an iPEPS [Fig. \ref{fig-PEPS} (a)] can be written as 
\begin{equation}
|\Psi\rangle = \text{tTr} \prod_{j} P^{[j]}_{s^j, a^j_ua^j_la^j_da^j_r} |s_j\rangle,
\label{eq-PEPS}
\end{equation}
where $\text{tTr}$ stands for the contraction on all shared indexes, $j$ runs over all lattice sites, and $|s_j\rangle$ denotes the local physical basis on the $j$-th site. The virtual indexes $\{a\}$ carry the entanglement of the iPEPS. In the thermodynamic limit, we introduce translational invariance, i.e., the local tensor $P^{[j]}$ satisfies $P^{[j]} = P$, without losing generality. The iPEPS representation can be used to construct non-trivial many-body states, as well as a variational ansatz of the ground state for a quantum Hamiltonian.

The inner product of an iPEPS with its conjugate defines a 2D TN as
\begin{equation}
 \langle \Psi | \Psi \rangle = \text{tTr} \prod_{j} T_{g^j_ug^j_lg^j_dg^j_r}.
\label{eq-TN}
\end{equation}
The local tensor in the TN satisfies
\begin{equation}
T_{g^j_ug^j_lg^j_dg^j_r} = \sum_{s^j} P_{s^j, a^j_ua^j_la^j_da^j_r} P^{\ast}_{s^j, a'^j_ua'^j_la'^j_da'^j_r},
\label{eq-LocalT}
\end{equation}
with $g^j_{\alpha} = (a^j_{\alpha},a'^j_{\alpha})$. See Figs. \ref{fig-PEPS} (b) and (c).

\begin{figure}[tbp]
	\includegraphics[angle=0,width=0.95\linewidth]{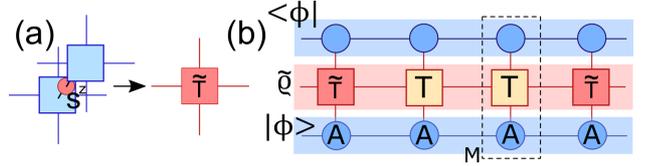}
	\caption{(Color online) (a) The graphic representation of the local tensor for calculating observables [Eq. (\ref{eq-ImpT})]. (b) The graphic representations of the boundary state $|\phi \rangle$ as an MPS formed by copies of the local tensor $A$ (blue shadow) and the MPO $\tilde{\varrho}$ formed by $T$ and $\tilde{T}$ (red shadow). The correlation function can be obtained simply by Eq. (\ref{eq-CorrelationMPS}). The transfer matrix $M$ [Eq. (\ref{eq-TMphi})] is given in the dashed square.}
	\label{fig-phi}
\end{figure}

Such a TN is very important as it contains fruitful information about physical properties of a quantum state. For instance, to calculate the correlation function
\begin{equation}
 \mathcal{C}(j_1,j_2) = \frac{\langle \Psi | \hat{S}^z(j_1) \hat{S}^z(j_2)| \Psi \rangle}{\langle \Psi | \Psi \rangle},
\label{eq-CorrelationF}
\end{equation}
one needs to contract a TN that is the same as $\langle \Psi | \Psi \rangle$ given by Eq. (\ref{eq-TN}) except for two tensors, each of which is obtained by
\begin{equation}
\tilde{T}_{g^j_ug^j_lg^j_dg^j_r} = \sum_{s^j s'^j} P_{s^j, a^j_ua^j_la^j_da^j_r} S^z_{s^js'^j} P^{\ast}_{s'^j, a'^j_ua'^j_la'^j_da'^j_r},
\label{eq-ImpT}
\end{equation}
with $S^z_{s^js'^j} = \langle s_j| \hat{S}^z |s'_j \rangle$ [Fig. \ref{fig-phi} (a)]. For simplicity, we assume these two operators locate in a same row of the lattice.

We may then introduce a 1D matrix product operator (MPO) \cite{MPO} $\varrho$ defined as an infinite tensor stripe in the TN [Fig. \ref{fig-phi} (b)]. Different from the previously proposed boundary Hamiltonian \cite{BoundaryH0}, the MPO $\varrho$ is actually the ``transfer matrix'' of the TN satisfying $\langle \Psi|\Psi \rangle = \lim_{N\to \infty} \text{Tr} (\varrho^N)$. It corresponds to a 1D effective quantum Hamiltonian defined in the space of the virtual bonds of the iPEPS. Then, by computing its dominant eigenvector $|\phi \rangle$ (dubbed as boundary state) of $\varrho$, one simply has
\begin{equation}
 \mathcal{C}(j_1,j_2) = \frac{\langle \phi |\tilde{\varrho}| \phi \rangle}{\langle \phi |\varrho| \phi \rangle},
\label{eq-CorrelationMPS}
\end{equation}
where $\tilde{\varrho}$ is obtained by replacing the $j_1$-th and $j_2$-th $T$'s in $\varrho$ with $\tilde{T}$. The calculations of other observables are similar. Note that $|\phi \rangle$ can be obtained using many different algorithms. Here, we choose the \textit{ab-initio} optimization principle of TN \cite{AOP}, where $|\phi \rangle$ is represented as a translationally invariant MPS formed by infinite copies of the local tensor $A$ as $\sum \cdots A_{g^j, a^j a^{j+1}} A_{g^{j+1}, a^{j+1} a^{j+2}} \cdots$ [see the blue shadow in Fig. \ref{fig-phi} (b)].

Let us further simplify Eq. (\ref{eq-CorrelationMPS}) by introducing the transfer matrix of $\langle \phi |\varrho| \phi \rangle$ [Fig. \ref{fig-phi} (b)] that reads
\begin{equation}
 M_{a^j g^l b^j, a^{j+1} g^r b^{j+1}} = \sum_{g^j g'^j} A_{g^j, a^j a^{j+1}} T_{g^j g^l g'^j g^r} A^{\ast}_{g'^j, b^j b^{j+1}}.
\label{eq-TMphi}
\end{equation}
Between two $\tilde{T}$'s, there exists the product of $|j_1-j_2|$ matrices $M$. Thus, one can readily see that the decay of the correlation function of the iPEPS versus the distance $|j_1-j_2|$ is dominated by the eigenvalue spectrum of $M$ in Eq. (\ref{eq-TMphi}), i.e., $\mathcal{C}(j_1,j_2) \sim (\frac{\Lambda_2}{\Lambda_1})^{|j_1-j_2|}$ with $\Lambda_i$ the $i$-th eigenvalue of $M$. Thus, the correlation length is given by $\Lambda_1$ and $\Lambda_2$ as
\begin{equation}
 % \xi = \frac{1}{\ln \Lambda_1 - \ln \Lambda_2}
 \xi = \frac{1}{\ln (\frac{\Lambda_1}{\Lambda_2})}
\label{eq-CorrelationLambda}
\end{equation}
Note that Eq. (\ref{eq-CorrelationLambda}) is independent of the specific choice of the correlation.

In fact, the correlation length of the MPS $|\phi \rangle$ is also given by Eq. (\ref{eq-CorrelationLambda}), implying that the criticality of an iPEPS can be determined by its boundary state $|\phi \rangle$. In the following, we take two 2D RVB states as examples and employ the scaling method of MPS \cite{EntCritic} to demonstrate the validity of our theory.

\textit{Resonating valence bond states on infinite two-dimensional lattices.}--- The iPEPS has many successful applications, one of which is to construct non-trivial many-body states. It has been shown that the nearest-neighbor RVB (NNRVB) state can be written in an iPEPS with the bond dimension $\chi=3$ \cite{PEPSCritical,RVBPEPS} ($\chi$ denotes the bond dimension of the iPEPS). Though the iPEPS is so simple, the physics is abundant and interesting. Such two NNRVB states possess the so-called topological order that cannot be characterized by any local parameters \cite{RVBPEPS}; NNRVB is gapped on kagom\'e lattice, but critical on a bipartite lattice \cite{RVBCritic}. We demonstrate that the boundary state can faithfully reproduce the criticality of the NNRVB states. 

% One of the simplest examples is the Z$_2$ topological state whose iPEPS has the bond dimension $\chi=2$.

\begin{figure}[tbp]
	\includegraphics[angle=0,width=1\linewidth]{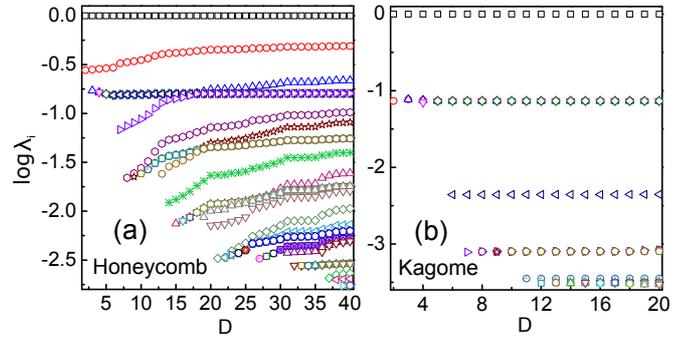}
	\caption{(Color online) The entanglement spectrum $\log \lambda_i$ of the boundary states of NNRVB on (a) honeycomb and (b) kagom\'e lattices versus the bond dimension $D$ of the boundary state. For honeycomb NNRVB that is critical, the values of the entanglement spectrum are squeezed when $D$ changes from 2 to 40. For kagom\'e NNRVB that is gapped as a comparison, the values of the entanglement spectrum stay unchanged with increasing $D$. For convenience, we normalize $\lambda$ so that $\lambda_1=1$.}
	\label{fig-SpcRVB}
\end{figure}

In Fig. \ref{fig-SpcRVB}, by varying the bond dimension $D$ of the boundary state MPS $| \phi\rangle$, we give the entanglement spectrum $\{\lambda_i\}$ of $| \phi\rangle$ for both NNRVB states. The spectrum $\{\lambda_i\}$ exhibits completely different patterns in critical or gapped situations. For the NNRVB on honeycomb lattice that is critical, the elements of each $\lambda_i$ are squeezed as the dimension $D$ increases. In comparison, the elements of $\lambda_i$ do not change with $D$ for the gapped NNRVB on kagom\'e lattice. Such results strongly indicate that the patterns given by gapped or critical iPEPS are essentially different from each other, providing a solid indicator to determine the criticality of the 2D iPEPS. It also means that for a gapped state, the $D$-largest Schmidt numbers can always be accurately determined with a finite $D$. For a gapless system, the criticality is encrypted in the scaling behavior of $S$ and $\xi$ against $D$. Note that an MPS with a finite $D$ always gives gapped state with an exponentially decaying correlation length [Eq. (\ref{eq-CorrelationLambda})]. 

\begin{figure}[tbp]
	\includegraphics[angle=0,width=1\linewidth]{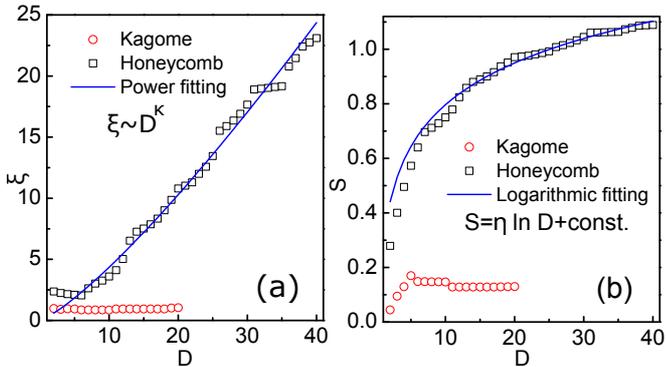}
	\caption{(Color online) (a) The correlation length $\xi$ and (b) the entanglement entropy $S$ against the bond dimension $D$ of the boundary state of NNRVB on kagom\'e and honeycomb lattices. For the honeycomb NNRVB, we find that $\xi$ increases in a power law as $\xi \sim D^{\kappa}$ with $\kappa \simeq 1.24$ and $S$ increases logarithmically as $S = \eta \ln D + \mathit{const}$. We have the central charge from Eq. (\ref{eq-CC}) as $c\simeq 1$, which corresponds to a free bosonic field in CFT. In contrast for honeycomb NNRVB, both $\xi$ and $S$ are small and converge to finite values as $D$ increases, suggesting a gapped boundary state.}
	\label{fig-XiEntRVB}
\end{figure}

Meanwhile, we find that the correlation length $\xi$ and the entanglement entropy $S=-\sum_i \lambda_i^2 \ln \lambda_i^2$ of the boundary states shows different scaling behavior against $D$. One can see from Fig. \ref{fig-XiEntRVB} that the boundary state of the honeycomb NNRVB is critical \cite{EntCritic}, satisfying
\begin{eqnarray}
 \xi &\sim& D^{\kappa}, \\
 S &=& \eta \ln D + \mathit{const.}
 \label{eq-XiEnt}
\end{eqnarray}
From CFT \cite{CFT,CFT_Ent}, the central charge characterizing the criticality of a 1D theory is defined as
\begin{equation}
 c=\frac{6 \eta}{\kappa}.
 \label{eq-CC}
\end{equation}
By fitting, we have $\eta \simeq 0.22$ and $\kappa \simeq 1.24$, thus the central charge $c =1.06 \simeq 1$, amazingly the same as the central charge obtained by Monte Carlo on the low-energy effective Hamiltonian of a 2D finite-size valence bond solid state \cite{cKitaev}. Our results show that the boundary state of the honeycomb NNRVB is described by a free bosonic field \cite{central1}.

For the boundary state of the kagom\'e NNRVB, both the correlation length $\xi$ and the entanglement entropy $S$ are small and converge as the bond dimension increases (Fig. \ref{fig-XiEntRVB}). These results suggest that the boundary state is gapped, which is consistent with the properties of the kagom\'e NNRVB.

\textit{Complexity of simulating critical ground states with tensor networks}--- We apply then our scheme to the variationally obtained iPEPS. To begin with, let us ask an important question: how large is the bond dimension of the iPEPS one needs to simulate the ground state at a critical point? For the MPS in 1D, the answer is infinite due to the logarithmic relation between the bond dimension and the entanglement entropy. Luckily, CFT allows us to access the criticality by the scaling with finite dimensions \cite{EntCritic}. For the iPEPS in 2D, there is not a simple \textit{yes-or-no} answer. Thanks to the network structure of iPEPS, the bond dimension does not have to be infinite to describe a critical state since the entanglement is carried by more than one bonds across the boundary of each subsystem. The NNRVB on honeycomb lattice is an example, which is critical, but given by an iPEPS with only $\chi=3$ \cite{PEPSCritical,RVBPEPS}.

If the iPEPS is obtained by a variational TN algorithm, this question becomes much more difficult to answer. First, the accuracy of the ground-state iPEPS is determined by not only the ansatz but also the optimization algorithms, for which there exist different variational strategies. Second, if the phases are gapped, one can expect an accurate location of the critical point between them because the iPEPS is believed to be faithful in both phases. But, it is still in debate whether the obtained iPEPS at the critical point truly captures the critical behaviour or not. For these reasons our scheme is of particularly great importance, since it enables us to efficiently identify the criticality for a given iPEPS.

We consider as an example  the XXZ model on honeycomb lattice, dewscribed by the Hamiltonian:
\begin{eqnarray}
 \hat{H} = \sum_{\langle i,j \rangle}  [J_{xy} (\hat{S}^x_i \hat{S}^x_j + \hat{S}^y_i \hat{S}^y_j) + J_z \hat{S}^z_i \hat{S}^z_j] + h \sum_i \hat{S}^z_i,
\label{eq-H}
\end{eqnarray}
where the summation is over all nearest neighbors and $h$ is the magnetic field in the $z$ direction. We choose $J_{xy}=0.5$ and $J_z=1$, and then, there exists an Ising-type quantum phase transition by changing the magnetic field. 

\begin{figure}[tbp]
	\includegraphics[angle=0,width=1\linewidth]{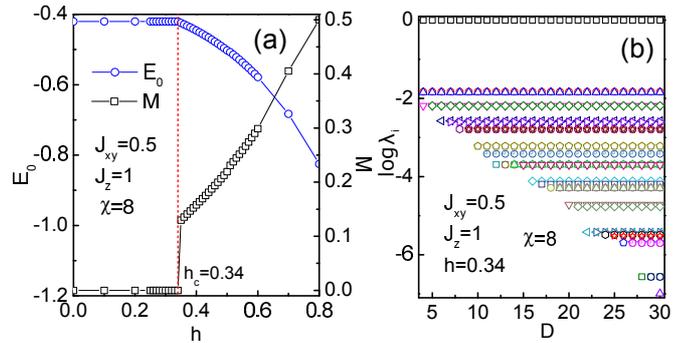}
	\caption{(Color online) (a) The ground state energy $E_0$ and magnetization $M$ versus magnetic field $h$. A quantum phase transition is found at $h_c=0.34$. (b) The entanglement spectrum $\log \lambda_i$ of the boundary state at $h=0.34$ do not move when the bond dimension $D$ increases, showing that the iPEPS is not critical. Here, we take the bond dimension cut-off of the iPEPS as $\chi=8$.}
	\label{fig-GSHC1}
\end{figure}

Applying the simple update algorithm \cite{TRG2}, we calculate the ground state energy $E_0$ and magnetization $M$. Here, we take the bond dimension cut-off of the iPEPS as $\chi=8$. A second-order quantum phase transition is clearly observed at $h_c=0.34$ [Fig. \ref{fig-GSHC1} (a)], indicated by the magnetization jump and the energy cusp. 

The determination of $h_c$ is quite reliable because the iPEPS can accurately give the states in both the anti-ferromagnetic and super-solid phases \cite{FrustrationBook}. Though, this does not mean the variational iPEPS at the critical point can really capture the criticality. In Fig. \ref{fig-GSHC1} (b), we show the entanglement spectrum $\log \lambda_i$ of the boundary state with different bond dimension $D$ (with $\chi=8$ fixed). When $D$ increases, the spectrum does not move or squeeze, which gives the same pattern as the gapped NNRVB on kagom\'e lattice. We also calculate the entanglement entropy and correlation length, both of which converge to finite values as $D$ increases. Besides, we change the magnetic field $h$ and the dimension cut-off of the iPEPS $\chi$, and observe no critical pattern of the entanglement spectrum.

Our results indicate the difficulties of obtaining the ground state at a critical point with the iPEPS approaches. Though the critical point can be accurately located, the iPEPS at the critical point is observed to be gapped, suggesting that the information of the criticality is lost. Surely, a lot  of issues remain open in this context, one of which is to test different TN algorithms to see how the optimization strategies take effects on capturing criticality. Another open issue is to explore the finite-dimensional TN representations of 2D critical quantum fields. Our scheme would be extremely useful to investigate these important issues.

\textit{Conclusion.}--- We proposed a robust scheme to determine the criticality of an infinite 2D quantum state with the help of iPEPS and TN. The entanglement spectrum $\{\lambda_i\}$ of the boundary state becomes squeezed as the the bond dimension $D$ increases for a critical iPEPS, and stays unchanged for a gapped one, giving two completely different $D$-$\lambda_i$ patterns. Our scheme is verified for the NNRVB states on kagom\'e and honeycomb lattices, where we find that the criticality of the honeycomb NNRVB is described by a $c=1$ CFT. Our work also unveils the difficulties of investigating the ground state at a critical point by variational iPEPS methods. 

With great versatility and flexibility, our scheme has a broad application on exploring the criticality of any other iPEPS's such as the string-net \cite{StringNet} and chiral \cite{PEPSChiral} iPEPS's, as well as those obtained variationally by any TN algorithms \cite{PEPSupdate}. Besides, it can also be used to investigate finite-temperature phase transitions with tensor product density operator algorithms \cite{ODTNS,NCD,FiniteTPEPS}.

% Future works can be done on identifying topological orders of 2D states by analyzing the entanglement of the degenerate states which are mixed in the MPO. This theory also gives implications on finite temperature calculations using the MPO-LTRG scheme as well as its higher-dimensional generalization \cite{ODTNS} on systems with spontaneous symmetry breaking and especially the topological properties. It is also interesting to find other universal scaling behaviors from different critical systems.

%\section*{Acknowledgements}
CP is grateful to financial support from UCAS for visiting ICFO and appreciates ICFO for hospitality. WL is indebted to Yun-Jing Liu for helpful discussions. This work was supported in part by the MOST of China (Grant No. 2013CB933401), the Strategic Priority Research Program of the Chinese Academy of Sciences (Grant No. XDB07010100), the NSFC (Grant No. 11474279 and 11504014), ERC AdG OSYRIS, Spanish MINECO (Severo Ochoa grant SEV-2015-0522, FOQUS grant FIS2013-46768), Catalan AGAUR SGR 874, Fundaci\'o Cellex, and EU FETPRO QUIC.

\end{document}